\documentclass[11pt]{article}
\usepackage{amssymb,amsmath, amsthm}
\usepackage{graphicx,picinpar,epsf}
\usepackage{float}
\usepackage[margin=2.5cm]{geometry}

\usepackage{subfigure}
\usepackage{mathrsfs}
\usepackage{color}
\usepackage{soul}
\usepackage{pdfpages}

\begin{document}

\title{From patterned response dependency to structured covariate dependency: categorical-pattern-matching.}
\author{\normalsize Hsieh Fushing\footnote{Correspondence: Hsieh Fushing, University of California at Davis, CA
95616. E-mail: fhsieh@ucdavis.edu}, Shan-Yu Liu\\
Department of Statistics, University of California, Davis,\\
Yin-Chen Hsieh\\
Department of Computer Science, University of California, Davis,\\
and Brenda McCowan,\\
School of Veterinary Medicine, University of California, Davis, U.S.A.}
\date{\today}

\maketitle

\section*{Abstract}
Data generated from a system of interest typically consists of measurements from an ensemble of subjects across multiple response and covariate features, and is naturally represented by one response-matrix against one covariate-matrix. Likely each of these two matrices simultaneously embraces heterogeneous data types: continuous, discrete and categorical. Here a matrix is used as a practical platform to ideally keep hidden dependency among/between subjects and features intact on its lattice. Response and covariate dependency is individually computed and expressed through mutliscale blocks via a newly developed computing paradigm named Data Mechanics. We propose a categorical pattern matching approach to establish causal linkages in a form of information flows from patterned response dependency to structured covariate dependency. The strength of an information flow is evaluated by applying the combinatorial information theory. This unified platform for system knowledge discovery is illustrated through five data sets. In each illustrative case, an information flow is demonstrated as an organization of discovered knowledge loci via emergent visible and readable heterogeneity. This unified approach fundamentally resolves many long standing issues, including statistical modeling, multiple response, renormalization and feature selections, in data analysis, but without involving man-made structures and distribution assumptions. The results reported here enhance the idea that linking patterns of response dependency to structures of covariate dependency is the true philosophical foundation underlying data-driven computing and learning in sciences.

\newpage
\section{Introduction}
Nearly all scientific researches are geared to acquire knowledge and understanding of systems of interest. So data generated from a system typically consists of measurements from an ensemble of subjects across possibly multiple response and many covariate features. It is known that, among these subjects, whether they are human, animal or plants or even cells, are likely interconnected, thus are all features interrelated on either response or covariate sides. Further the interconnections and interrelatedness weave interacting relations between subject and features on either response or covariate sides. Hence, as a rule, each system data set contains these three fronts of dependency with unknown structures.  Even through their existences are known and recognized as essential parts of most scientific systems, such structural dependency is hardly taken into account in statistical modelings or supervised machine learning. Therefore these three forms of dependency are completely left out in the resultant analysis. This phenomenon is not unexpected. Since modeling a system is task of ``synthesis''. It is all but impossible, as Physics Nobel laureate P. W. Anderson (1972) had pointed out more than four decades ago.

Thus, instead of attempting the impossible task of synthesizing all distinct scale-specific dynamics into one model system, we should at least take a humble stand by honestly exploring what knowledge and understanding can a system data set offer. From this stand point, the goal of systemic endeavors in general can be generically prescribed as linking the response's computable authentic structural dependency to the covariate's. To be able to possibly achieve such a goal, it is natural as well as necessary to build such a linkage between a response's data matrix and a covariate's data matrix because each matrix representation is expected to keep its hidden idiosyncratic structural dependency intact. As for extracting the two dependency structures respectively and then making the linkage, the employed computational and learning algorithms are preferably involving with only simple mechanistic arrangements upon the original data, that is, no unrealistic statistical assumptions being attached. Furthermore, since such linkages embrace authentic system understandings and knowledge, the computational results are better literally readable and pictorially visible. In this paper we demonstrate and illustrate such an achievable goal through five real data examples.

Traditional popular statistical modeling methodologies, such as analysis of variance (ANOVA), Logistic or linear regression models, and even the recently popular supervised machine learning algorithms ignore structural dependency entirely. Such ignorance can be seen through the following three simple facts. First fact is that nearly all these popular existing techniques can only accommodate {\it one} single response feature. Second fact is that the conditioning argument is applied to completely detach potential inferences from the covariate's structural dependency. The third fact is that, on top of imposing independence among subjects, the employed models have to rely on built in assumptions of linearity and homogeneity. All these assumptions are often impossible to meet. That is to say that users of these models are destroying structural dependencies of responses and covariates in order to stuff observed data and assumed structures through the likelihood funnel into parameterized mathematical formulations. Such a data-analysis resulted from a process like sausage-making is hardly authentic because, like a sausage, it is a mix of natural as well as man-made ingredients.

Further these model-based techniques can only accommodate selective data types, in others words, a data type often is the decisive factor in choice of models. For instance, when the one-dimensional(1D) response feature is binary, the logistic regression model is the choice. This logistic modeling framework breaks down when the response feature has more two categories. Then, when the 1D response feature has more then two categories, then multivariate Analysis of Variance (MNOVA) is to be used. For MNOVA, multivariate Normality and independence assumptions have to assume upon the measurements of covariate features. That is, only continuous data type of covariate feature is feasible in MNOVA.

Next when the 1D response feature is continuous, then a linear regression or generalized linear model has to be used. If a qualitative categorical feature is present in the covariate side, then this feature has to be always forcefully transformed into multiple surrogate binary variables. This transformation is purely for technical reasons. Since a likelihood function or matrix is not operational with categorical entries. For instance, a categorical feature having 10 categories are to be represented by 9 separate binary features. These 9 binary variables indeed create confusion in their dubious interpretations. However, when there are more than one features on the response side, all aforementioned modeling structures break down. This is the so-called multiple response issue. So far there exist no satisfactory resolutions in literature after it was raised more than half century by John Tukey (1962).

At this era of big data, the effects of the above known shortcomings of statistical modeling and inferences would certainly get more serious than ever before. So it is appealing now to practically avoid extracted information being compromised by hypothesized structures and its contents being by and large dictated by data types. The key steps for carrying out this appeal are: first, to give up all man-made assumptions; secondly, to focus on authentic structural dependency on response and covariate sides; thirdly, to make the linkages between these two dependency structures. We envision that an unified platform is possible to coherently implement the three steps toward system knowledge and understanding. In this paper, such a platform is proposed based on a newly developed computing paradigm Data Mechanics and a fundamental categorical-pattern-matching approach.

To begin with our developments, we take a rather unconventional approach: {\bf basically re-normalizing all features into digital-categorical ones}. We believe that digital-categorical is the most fundamental data type. The categorical nature will make possible for employing combinatorial information theory, which provides the most basic and reliable evaluation of associative relation. It relies basically on counting only. The conditional-entropy deduced from the combinatorial information theory can accommodate nonlinear relational associations and is meaningful with no need of any hypothetical assumptions. In contrast, the correlation relies on linearity and Normality assumptions, so is not robust to diverse waveforms and perturbations away from these assumptions.

We specifically focus the re-normalization protocol on: 1) how a continuous feature is re-normalized into a digital-categorical via its possibly-gapped histogram; 2) how discrete and qualitative features are also re-normalized with respective to their closely associated digital-categorical features. It is important to note that the purpose of re-normalization is to make all features digitally comparable. Such comparability is the foundation for computing and representing structural dependency on response and covariate sides.

The computing for structural dependency is primarily performed through the application of Data Mechanics, an unsupervised learning algorithm developed in Fushing and Chen (2014) and Fushing, et al. (2015), which merely carries out permutations on row- and column-axes of a matrix in order to achieve the nearly minimum total-variation, or energy, on the matrix-lattice. By applying Data Mechanics on the matrix of mutual conditional-entropy of response' features, its permuted matrix with lowest energy would reveal an overall skeleton of response's structural dependency, which typically shows visible multiscale block patterns through its heatmap. Here we introduce and identify the synergistic feature-groups, each of which is a potential system mechanism. Likewise covariate's structural dependency and its synergistic feature re-grouping is computed and identified.

Upon a data sub-matrix pertaining to a synergistic feature group, the Data Mechanics is applied again to frame such a sub-matrix by two clustering trees superimposed onto its row- and column-axes. Then a mechanism-specific structural dependency is revealed through the multiscale block-patterns on its heatmap. To link one response mechanism to one covariate mechanism, we propose to make the linkage from the response's clustering tree on subjects (row-axis) to the covariate's clustering tree on subjects. The strength of such a linkage is again evaluated via the directed conditional-entropy based on the combinatorial information theory. Confirmations of such linkages and evaluations of predictive error-rates are simply performed through the same theory. Such a procedure of linkage building is called categorical-pattern matching. Thus such matrix-based causal linkage doesn't need or involve any man-made structure or assumption.

\section{Reviews of conceptual foundations in Data Analysis}

\subsection{Evaluation of amount of information conveyed by $X$ with regard to $Y$.}
In his 1965 paper with title ``Three approaches to the quantitative definition of information,'' A. N. Kolmogorov said that\\
``.. It is only important for me to show that the mathematical problems associated with a purely combinatorial approach to the measure of information are not limited to trivialities.''

Indeed the combinatorial approach has been well known in Information Theory since  C. E. Shannon's (1949, 1951) pioneer works. However, outside of Communication and Information Theories, its use in real world data analysis is not yet evident, neither popular nor widespread. This is not because it is not useful, but rather because it has been overshadowed by the concept of ``correlation'' in Statistics, which is nothing but an inner product of two unite vectors in mathematics without the rigorous checking on the bivariate Normality assumption.

In this paper we discuss that the combinatorial approach of information in fact gives us an universal measure of associative relation between two variables. Based on conditional entropy, this relational association concept will be seen as especially suitable for unsupervised machine learning and its inferences, in which the ``sample-to-population'' sense is not involved. Along the developing process, we also reflect why the linearity backbone of correlation can cause invalid and even often misleading interpretations on real data. Before introducing such an measure of entropy based associative relation, it is beneficial to review this combinatorial approach of Information.

Consider and denote the amount of uncertainty, say $A(N)=UC(\frac{1}{N},....., \frac{1}{N})$, of choosing one subject with uniformly equal potentials among $N=m\times n$ subjects contained within an ensemble. If these $N$ subjects are divided in $m$ sub-ensembles of size $n$, then the equal-potential sampling scheme on $N$ subjects is equivalent to first sampling with equal potentials from the collection of $m$ sub-ensembles, and secondly sampling one subject with equal potentials from the chosen sub-ensemble of $n$ subjects. This so called composition rule, see Janyes(1957), implies that the uncertainty $A(N)=A(n\times m)=A(m)+A(n)$. Shannon (1948) has determined that such $A(N)=C\times log{N}$ up to a constant $C$. Let's choose a $C$, such that
\[
A(N)=-\sum^N_1 \log_2{\frac{1}{N}}=\log_2{N}.
\]

In general, if $N$ subjects are number marked and partitioned in $K$ color-coded sub-ensembles possibly unequal sizes $(N_1, ...N_K)$, that is, two variables are defined upon this ensemble of $N$ subjects: let the variable $Y$ be the number-coding from 1 to $N$, and $X$ be the color-coding. Let the $K$ color-coded sub-ensembles have their proportion being denoted as $\{p_k=\frac{N_k}{N}\}^K_{k=1}$, then the entropy of discrete variable $X$ pertaining to sampling with probability $\{P_k\}^K_{k=1}$ is generically calculated and denoted as
\[
H(X)=-\sum^K_{k=1} p_k \log{p_k}=A(N)-\sum^K_{k=1}p_k A(N_k).
\]
This equation says that the amount of information conveyed by variable $X$ with regard to the variable $Y$, say $I[Y:X]$, is exactly equal to $H(X)$, see Kolmogorov(1965).
Here we use the notation for the above equation as
\[
E[Y \rightarrow X]=H(Y)-H(Y|X)=H(Y)-\sum^K_{k=1} p_k H(Y|X=k).
\]

Based on such combinatorial information theory, the mutual conditional-entropy for two clustering compositions is pictorially illustrated in Box1, while their formulas are contained in Box2. Specifically the tree at bottom of Box1 is a subset of X-covariate tree for $Y$(color coding) given $X=a$ and its corresponding conditional entropy is developed at the beginning of Box2.

\includepdf[pages={1}]{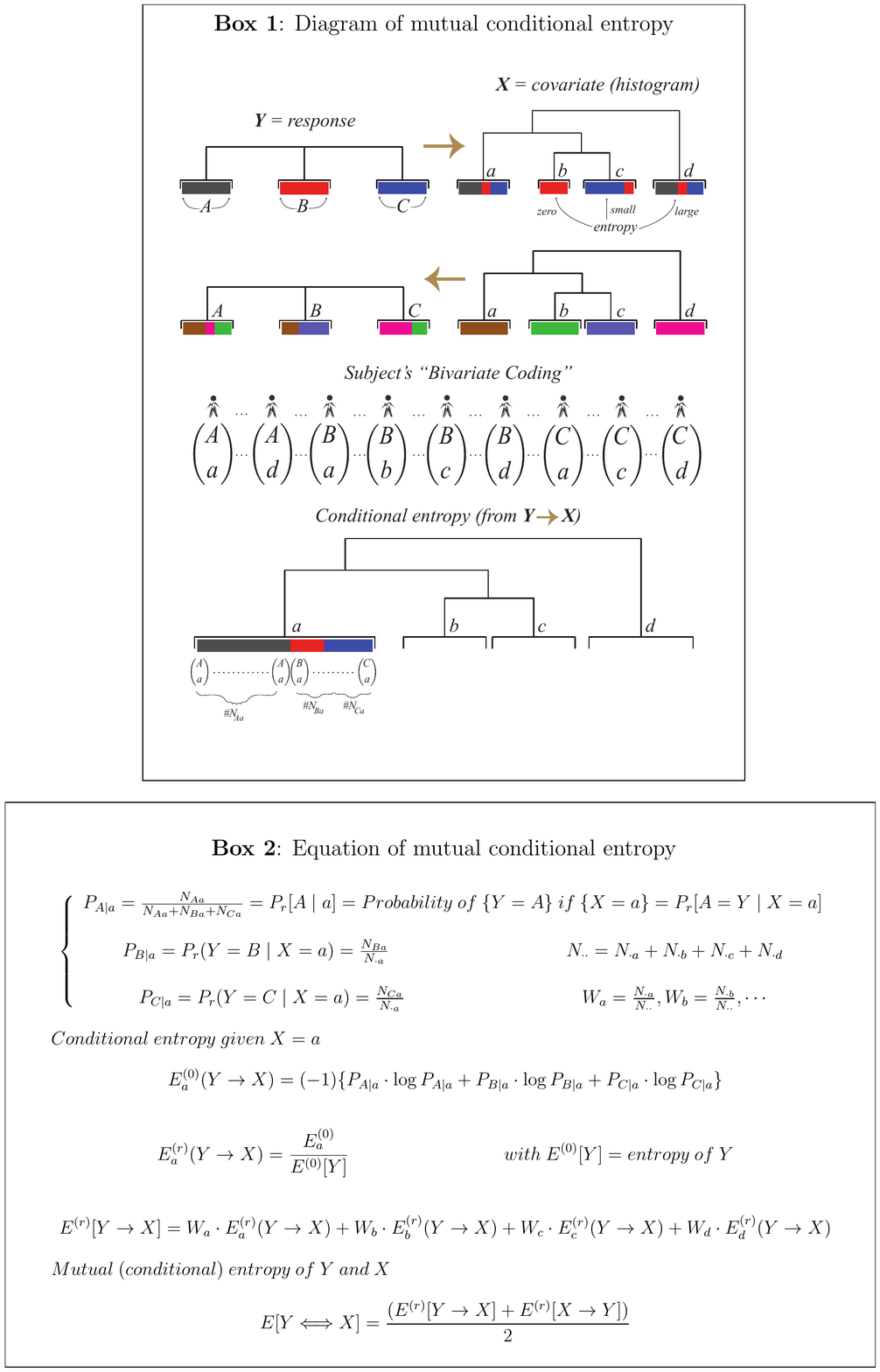}


\subsection{Factors and mechanisms in a system of interest.}
When a system is under study, it is universal that many dimensions of feature-specific measurements are observed or measured from subjects, which are constituents of the system. Unless they are coordinated according to temporal, spatial or other known axes, these features are typically unorganized with respect to a known framework. Nonetheless, just as coordinated features likely manifest evolving system states along the axes, these unorganized features also likely comprised of various distinct mechanisms of dependency. As such, system states and mechanisms as system's major components are popularly called factors in many scientific fields, especially in psychology and economics and many other social sciences.

The popularity of factor analysis is based more on computational convenience than on meaningful interpretations, see Tukey (1962). Specifically these factors are conveniently computed via principle component analysis (PCA), singular value decomposition (SVD) and their dynamic variants based on covariance matrices. Hence factor analysis is mainly used as a way of achieving linearity based dimension reduction and retaining major proportion of information contents under normality. However, since intricate patterns of dependency among features potentially go far beyond dyadic correlations, such factor analysis often incurs information loss and unnatural representations of underlying system mechanisms. That was partly why Tukey (1962) strongly discouraged applications of factor analysis.

In order to naturally and explicitly reveal system mechanisms, it becomes necessary to demonstrate structural dependency among all features included in the data. Given that distinct mechanisms involve with distinct feature-groups of different sizes, the issue of how to re-group features to show distinct mechanisms becomes a pressing issue in any system study. One universal concept of dependency considered here is based on $E[Y \rightarrow X]$ and $E[X \rightarrow Y]$ of two features denoted by $X$ and  $Y$, that is, if $X$ is capable of conveying a non-negligible amount of information in relation to $Y$, or vice versa, then $X$ and $Y$ are dependent. After building a mutual conditional-entropy matrix, subsequently, as will be demonstrated in sections blow, an unsupervised learning algorithm is applied to perform the task of feature regrouping. That is, a synergistic group of features is seen as constituting a mechanism, while two synergistic groups being antagonistic would be seen as two separate mechanisms. Unlike the logistic and linear regression models in statistics, our proposed computational protocol will link one synergistic response-feature group to one or several synergistic covariate-feature groups. This proposal interestingly fulfills Tukey's (1962) postulation of appealing to Taxonomy and classification methodologies on the multiple response issue.

\subsection{System knowledge}
In a system study, the primary goal of data analysis is to compute and organize visible knowledge loci pertaining to the linkages from a response's mechanism to covariate's mechanisms. Since all system's mechanisms on both sides share the common space of constituent subjects belonging to the system under study. The linkages are supposed to be seen and built through this common space of subjects.

To be more specific, the simplest, but most important form of system knowledge linkage is prescribed with {\bf heterogeneity} as: One serial uniform pattern-blocks framed by a serial synergistic covariate-feature groups and a cluster of subjects nearly exclusively explain one part of one whole pattern-block framed by a synergistic response-feature group and a {\bf larger} cluster of subjects.

Here a block-pattern is taken as a knowledge locus, and a linkage via the exclusiveness is meant to be equipped with an extreme conditional entropy ($E[Y \rightarrow X]$) of being near zero. Via this heterogeneity, scientists specifically figure out how the measurements of a synergistic response-feature group upon a subject-cluster can be explained collectively by multiple distinct series of block-patterns manifested by serial synergistic covariate-feature groups upon a partition of the original subject-cluster. From this perspective, our data analysis is clearly fundamentally distinct from convention statistical modeling and supervised learning methods, which heavily rely on the hypothesized sample-to-population homogeneity.

Unsupervised learning paradigms and combinatorial approach, as will discussed in the Method section, are ideal for computing and discovering knowledge representations and achieving the constructions of such linkages. In fact this computational paradigm and representational approach provide a means to safe guard against such apparent dangers of man-made inconsistency and fallacy. That is, system knowledge derived from this approach will necessarily reveal patterns of multiple scales that are realistically available from data.

\section{Methods}
\paragraph{$\S$[Motivations and goals]}
To motivate and introduce the computational paradigm employed for extracting system knowledge loci, we begin by explaining why the popular Logistic regression model in statistics is not expandable from mathematical perspective. It is worth emphasizing that similar explanations would be applicable to the linear regression model as well. A non-classical view of a Logistic regression model is expressed in Fig.1(a) with two horizontal half-lines being designated for the binary response categories: $Y=1$ and $Y=0$, while linearly combinations of covariates $\beta X$ are correspondingly marked on these two half-lines. With respect to this display, the optimal $\beta$ is seen to achieve the least overlapping between the two ranges: $R_e[1]=[\min{\{\beta X_i|Y_i=1\}},\max{\{\beta X_i|Y_i=1\}}]$ and $R_e[0]=[\min{\{\beta X_{i'}|Y_{i'}=0\}},\max{\{\beta X_{i'}|Y_{i'}=0\}}]$. This non-classical view is indeed fundamental because its expanded version of display can accommodate the setting of response variable $Y$ having more than two categories. With such a fundamental view, why the Logistic regression model is still not expandable? The answer essentially lies with the simple fact that even the straight forward overlapping evaluation among all induced ranges can not afford a single smooth functional form of $\beta$.

\begin{figure}[H]
  \centering
  \includegraphics[width=5in]{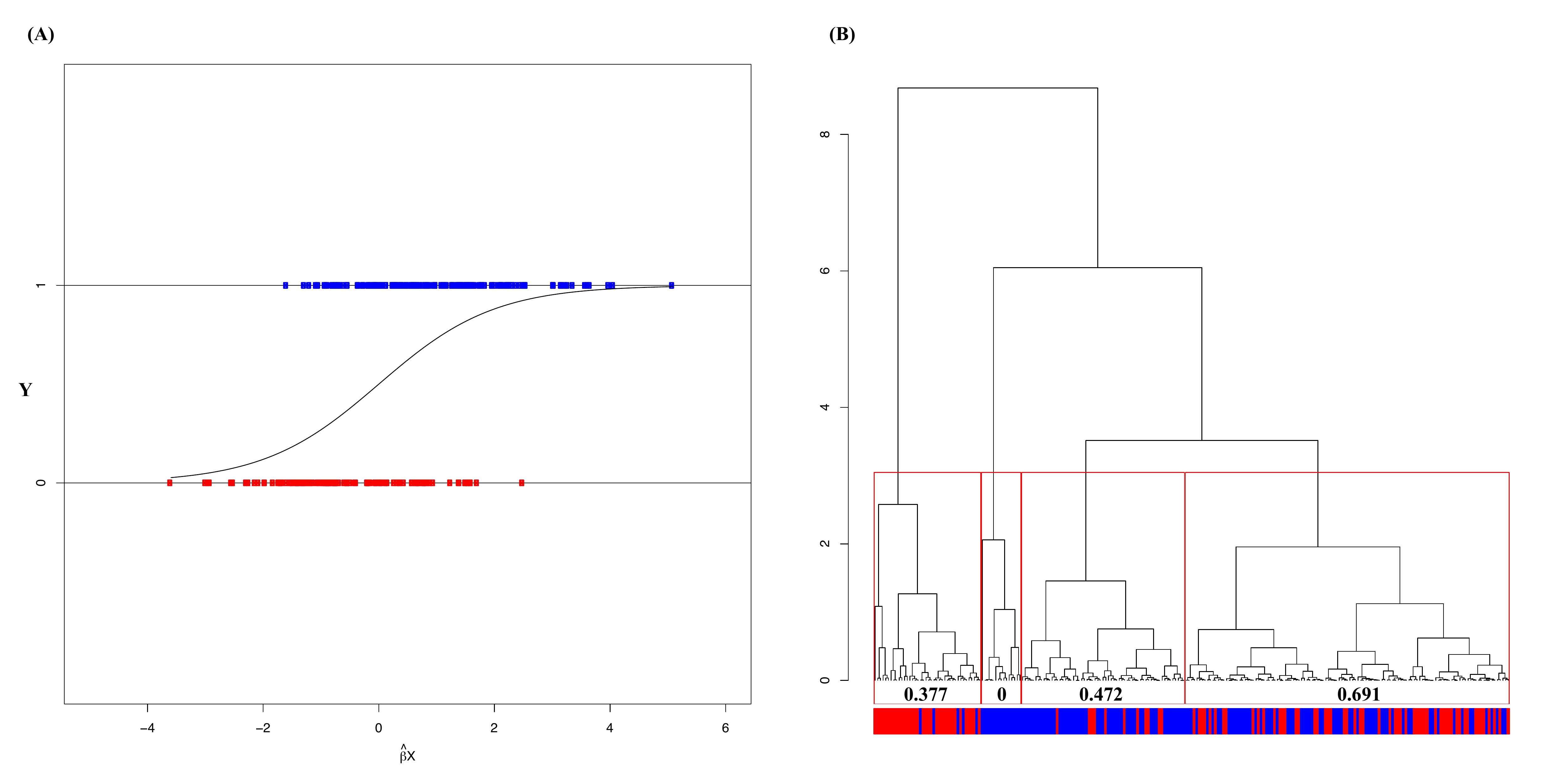}

  \caption{An expandable Logistic regression setup and possibly heterogeneity. (A) Binary horizontal layout with respect to $\hat{\beta}X_i$ with MLE $\hat{\beta}$.; (B) Histogram of $\hat{\beta}X_i$ with calculated entropies for each cluster.  A high degree of overlapping between the two horizontal layout in (A) indicates inefficiency of Logistic regression. In contrast the heterogeneity  within each gender categories in (B) gives rise to precise results in clusters of $\hat{\beta}X_i$ with low entropies.}
\end{figure}

Apart from being not able to accommodate a response variable beyond binary response variable, a Logistic regression also critically suffers from its linearity imposed constraint of homogeneity. It can not accommodate heterogeneity such as shown in the Fig.1(b). The hierarchical clustering tree of $\hat{\beta}^{(MLE)}X$ at its 4-cluster tree-level is capable of revealing heterogeneous information contents. That is, instead of counting the overlapping $R_e[1]$ and $R_e[0]$, indeed we can extract more information by breaking the range $R_e[1]\bigcup R_e[0]$ into pieces in a natural way. Informative patterns are observed upon these four clusters (from the left to the right) as follow:1) primarily dominant by Red color-coded subjects ($Y=0$); 2) purely Blue color-coded subjects ($Y=1$); 3) primarily dominant by Blue color-coded subjects ($Y=1$); 4) a mixture of Red and Blue color-coded subjects. It is surprising that by allowing heterogeneity, the misclassification result of a Logistic regression with a given threshold can be very much improved and more precisely understood. This hierarchical clustering tree provides an extra advantage that there is no need to choose an ad hoc threshold to count for the false-positive (FP) and false-negative(FN).

However the heterogeneity through the hierarchical clustering tree of $\hat{\beta}^{(MLE)}X$ provides only one limited aspect of intrinsic heterogeneity contained within the data because of being limited by one specific direction of covariate features pertaining to $\hat{\beta}^{(MLE)}$. Hence it is realistic to expect that, if data's whole intrinsic heterogeneity is properly computed and suitably represented and visualized, then the full information contents contained within data should be seen. Here such intrinsic heterogeneity is taken as system knowledge. In order to reveal such heterogeneity fully, as the ultimate goal of our data driven computations in this paper, we advocated unsupervised learning and computing paradigms, as would be briefly described below. It is worth emphasizing that the importance and essence of such paradigms is to make all computed results free from man-made constraints or distortions via invalid modeling assumptions. This important and essential point is particularly relevant to data analysis in the age of big data.

\paragraph{$\S$[Possibly-gapped histogram based re-normalization]}
Let ${\cal M}_0$ be an observed $n \times m$ data matrix with $n$ subjects being arranged along the row-axis and $m$ features along the column-axis. Each feature specific column has to undergo a digital re-normalization procedure based on a data-driven possibly-gapped histogram as illustrated in Fig.2, see detailed computations and algorithmic programs in Fushing and Roy (2017). Such a possibly-gapped histogram based re-normalization is designed to achieve three goals making:1) all columns free from their idiosyncratic measurement scales; 2) all features' ranges comparable; 3) digital coding naturally reflecting the 1D data-structural geometry. In summary, a feature's possibly-gapped histogram has to effectively approximate its empirical distribution function, which may have horizontal gaps. That is, no continuity assumption is implicitly imposed here. By doing such a normalization, all features involved could possibly be able to contribute nearly equally to the similarity or distance measurements of all feature-pairs as well as all subject pairs.

However, when the $m$ features are mixed in data-types: continuous, discrete and categorical, as would be seen in the Heart data below, the re-normalization becomes a rather tricky issue to be resolved in order to achieve a large degree of uniformity. Here we suggest a guideline: {\bf two features with relatively low mutual conditional-entropy should be similarly digitally coded}.
Denote the $n \times m$ re-normalized data matrix be ${\cal M}_{\circledR}$.

\begin{figure}[H]
  \centering
  \includegraphics[width=5in]{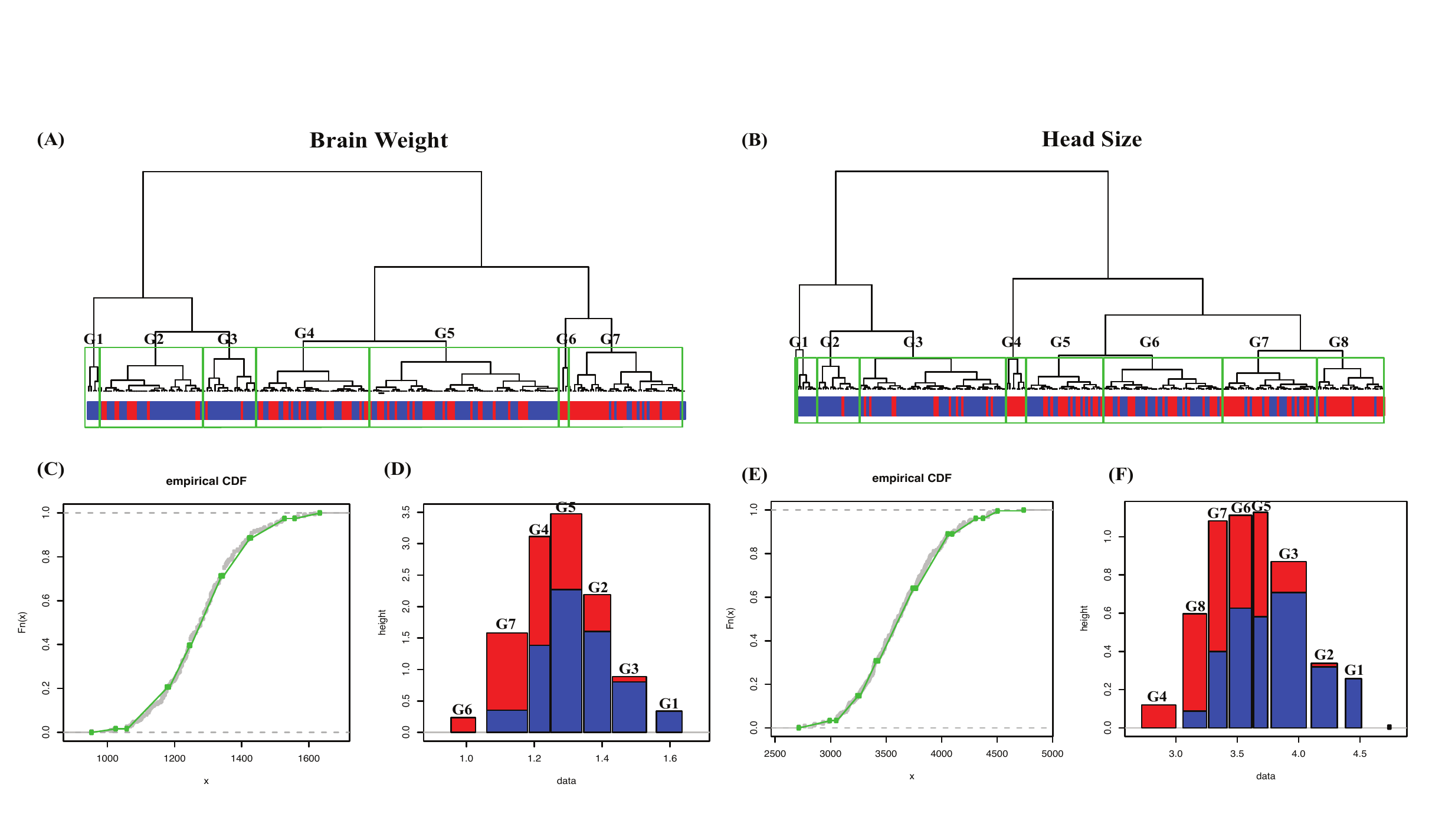}

  \caption{Two features' hierarchical clustering trees and corresponding empirical distributions and possibly-gapped histograms. (A)Head size's hierarchical clustering tree marked with 8 clusters;(B) Brain weight's hierarchical clustering tree marked with 7 clusters (C)The empirical distribution of head size superimposed with a 8-piece linear approximations showing with possibly-gaps; (D) The possibly-gapped histogram with 8 bins colored with gender proportions. (E)The empirical distribution of brain weight superimposed with a 7-piece linear approximations showing with possibly-gaps; (F) The possibly-gapped histogram with 7 bins colored with gender proportions. It is noted that the both histograms in (D) and (F) have two visible gaps separating the far-left and far-right bins. This is the strong evidence of dependency between these two features.}
\end{figure}

\paragraph{$\S$[Synergistic-vs-antagonistic feature grouping via Data Cloud Geometry (DCG) algorithm.]}
Upon this $n \times m$ re-normalized data matrix ${\cal M}_{\circledR}$, we can compute a $m \times m$ mutual conditional-entropy matrix, say $Xi$, for all feature pairs $(Y, X)$ as follows:
\[
2E[Y \Leftrightarrow X]=\frac{H(Y|X)}{H(Y)}+ \frac{H(X|Y)}{H(X)}=E[Y \Rightarrow X]+E[X \Rightarrow Y],
\]
where $(Y, X)$ is generic bivariate digital coding, i.e. two separate columns of ${\cal M}_{\circledR}$, pertaining two features in the feature node-space.

Then the $m \times m$ mutual conditional-entropy matrix $\Xi$ can be taken as a distance matrix for feature-grouping computations. The Data-Cloud-Geometry (DCG) computing algorithm employed here aims at building an Ultrametric clustering tree, say ${\cal T}[{\Xi}]$. The key concept underlying DCG algorithm is to discover multiple essential scales, to which clustering relational patterns are evident. As such natural clustering compositions must be scale-dependent and need to be discovered. When they are synthesized with respect to decreasing identified scales, an Ultrametric clustering tree is built with one scale corresponding to one tree level. See detailed algorithmic computing in Fushing and McAssey (2010) and Fushing et al. (2014).

\includepdf[pages={1}]{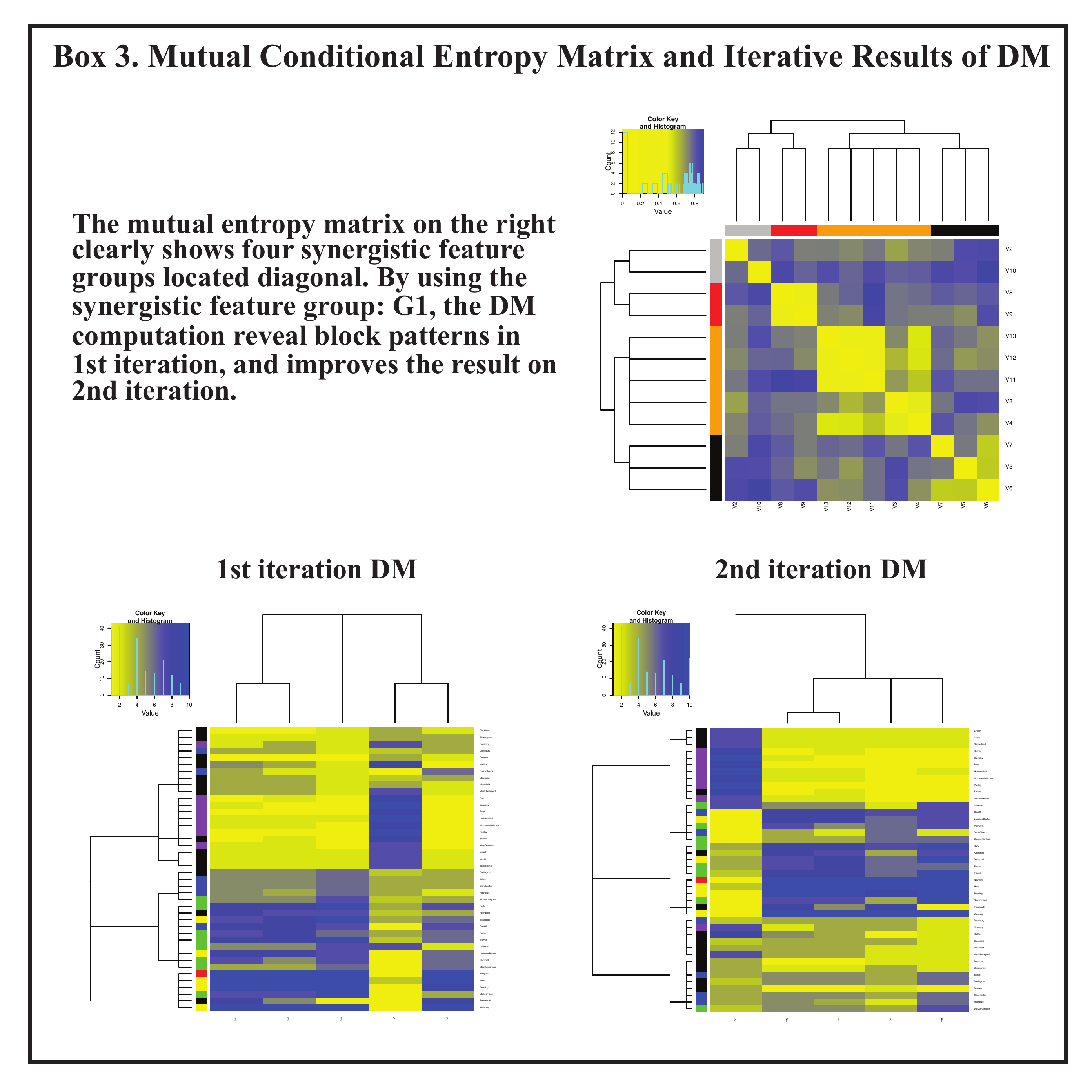}

By superimposing this Ultrametric tree on row and column axes of $\Xi$, its framed matrix lattice will naturally show multiscale block-patterns, as illustrated in panel(A) of Box3 with four colored coded synergistic feature groups. The blocks on diagonal of various sizes are blocks consisting of relative small mutual conditional-entropies, so they are synergistic feature-groups with various degrees. In contrast off-diagonal blocks having large mutual conditional-entropies indicate antagonistic relations between feature-groups. In summary the chief merit of mapping out synergistic feature-groups against antagonistic ones is to make complex nonlinear dependency structures visible.

\paragraph{$\S$[Data mechanics on matrix data]}
Features sharing a synergistic feature-group are highly dependent because they potentially constitute one distinct mechanism. Such dependency will therefore allow unsupervised learning algorithms to more effectively reveal fine scale interacting relational patterns between subject-clusters and feature-clusters. That is, the row-axis of ${\cal M}_{\circledR}$ should be partitioned according to the hierarchy of ${\cal T}[{\Xi}]$, so that different involving mechanisms are discovered and visualized, as seen in panel(A) of Box3.

The unsupervised learning algorithm employed here is called Data Mechanics (DM), see Fushing and Chen (2014) and Fushing et al. (2015). Data Mechanics is an iterative algorithm that build one DCG-based Ultrametric tree on row-axis and one on column-axis alternatingly. In each iteration, the distance metric is updated with respect to the tree structure on the previous axis. Denote the final two Ultrametric trees ${\cal T}[{\cal M}_{\circledR}]_R$ and ${\cal T}[{\cal M}_{\circledR}]_C$ on row- and column-axes, respectively. The overall goal of DM computing is to permute rows and columns such that multiscale blocks framed by the two marginal trees ${\cal T}[{\cal M}_{\circledR}]_R$ and ${\cal T}[{\cal M}_{\circledR}]_C$ are as uniform as possible.

The uniformity within each of the finest scale block collectively forms the stochastic structures contained within ${\cal M}_{\circledR}$, while the multiscale blocks framed by two marginal trees ${\cal T}[{\cal M}_{\circledR}]_R$ and ${\cal T}[{\cal M}_{\circledR}]_C$ constitutes the deterministic structures contained within ${\cal M}_{\circledR}$. These two coupled structural components are taken to be the information content and termed coupling geometry of ${\cal M}_{\circledR}$. The DM and its coupling geometry are illustrated through panels (B-D) of Box3 with three heatmaps corresponding to three iterations, respectively. The heatmap from 1st iteration of DM is apparently very much improved by that of 2nd and 3rd iterations. The later two are exactly the same. This fact indicates that the number of iterations needed is in general small.

\paragraph{$\S$[Organization of knowledge loci into information flows]}
Now consider a $n \times m_{Re}$ response data matrix, ${\cal M}^{(Re)}_0$, and one $n \times m_{Co}$ covariate data matrix ${\cal M}^{(Co)}_0$. The ultimate computing goals are to identify all essential system mechanisms involving both response and covariate sides, and then discover all related knowledge through associative patterns that link a response mechanism to a serial covariate mechanisms. The computations for achieving such goals are carried out in the following steps.

\noindent
{\bf [Algorithmic steps for discovering and confirming information flows:]}\\
\begin{description}
\item[1] {\bf [Re-normalizing all features]:} Matrices ${\cal M}^{(Re)}_0$ and ${\cal M}^{(Co)}_0$ will undergo their column-by-column re-normalization, as described in the above paragraph. The resultant digital-coding matrices are denoted as ${\cal M}^{(Re)}_{\circledR}$ and ${\cal M}^{(Co)}_{\circledR}$, respectively.

\item[2] {\bf[Re-grouping synergistic features]:}Upon ${\cal M}^{(Re)}_{\circledR}$, its $m_{Re} \times m_{Re}$ mutual conditional-entropy matrix $\Xi_{Re}$ is computed, so is $m_{Co} \times m_{Co}$ mutual conditional-entropy matrix $\Xi_{Co}$ based on ${\cal M}^{(Co)}_{\circledR}$. Then essential mechanisms on response and covariate sides are identified through DCG-based Ultrametric clustering trees ${\cal T}[{\Xi_{Re}}]$ and  ${\cal T}[{\Xi_{Co}}]$, respectively. Hence synergistic feature-groups on response and covariate sides are collected respectively.

\item[3] {\bf[Discovering block-patterns via Data Mechanics]:}Each data submatrix of ${\cal M}_{\circledR,i}$ corresponding to each synergistic feature-group would undergo Data Mechanics computations, and its row-marginal tree ${\cal T}[{\cal M}_{\circledR,i}]_R$ is collected.

\item[4] {\bf[Exploring information flows via Categorical-pattern matching]:} Each row-marginal tree ${\cal T}[{\cal M}^{(Re)}_{\circledR,i}]_R$ on the response side will be paired with each every row-marginal tree ${\cal T}[{\cal M}^{(Co)}_{\circledR,j}]_R$ on the covariate side, and compute the directed response-to-covariate conditional entropy as $E[Y \Rightarrow X]$. A low value of such directed conditional entropy implies that there exist strong associative patterns. Specifically a strong associative pattern is identified as a cluster or branch of ${\cal T}[{\cal M}^{(Co)}_{\circledR,j}]_R$ being nearly exclusively belonging to one particular cluster or branch of  ${\cal T}[{\cal M}^{(Re)}_{\circledR,i}]_R$. Such a associative pattern is termed a knowledge locus.

\item[5] {\bf [Information flows]:} Organize all knowledge loci with respect to a series of coupling geometries via a series of heatmaps. This is representation is called Information flow of one response mechanism to a serial covariate mechanisms.

\end{description}

\paragraph{$\S$[Confirming an information flow and calculating its error rate.]}
Due to the exploratory nature of an organization of knowledge loci, a result of categorical-pattern matching, as in the Step 4 {\bf[Exploring information flows via Categorical-pattern matching]:}, needs a formal confirmation, and then its error rate has to be evaluated. All these computations are scale-dependent, but rather simple and elementary. Here the scale-dependence is referring to a clustering composition of subjects pertaining to a chosen tree level of the row-marginal tree ${\cal T}[{\cal M}^{(Re)}_{\circledR,i}]_R$ coupled with a clustering composition of subjects pertaining to a chosen tree level of row-marginal tree ${\cal T}[{\cal M}^{(Co)}_{\circledR,j}]_R$.

Given a pair of clustering compositions respectively coming from response and covariate sides, observed directed conditional entropies are evaluated on each individual cluster of the clustering composition on the covariate side, as illustrated in Box 2. To confirm any pattern contained in an information flow, we apply the simple random sampling without replacement (with respect to the proportions of subjects in the clustering composition on the response side) to calculate a simulated entropy distribution and accordingly the P-values with respect each observed entropy.

For an error rate of an information flow with only one covariate synergistic feature-group, either majority rule or randomized rule can be applied within each cluster on the covariate side to calculate individual cluster's error rates, and then a weighted overall version is calculated with respect to sizes proportions of the covariate clustering composition. The reason underlying this simplicity is given as follows. Randomly select one subject and remove its cluster membership on the response side. The key is that based on the semi-unsupervised learning paradigm, this subject's row covariate vector needs to participate in the construction of row-marginal tree ${\cal T}[{\cal M}^{(Co)}_{\circledR,j}]_R$. Thus this row-marginal tree ${\cal T}[{\cal M}^{(Co)}_{\circledR,j}]_R$ is invariant with respect to any random choice of subject. That is, the selected subject keeps its original position in the clustering composition on the covariate side. Therefore, by repeating this random selection for a large number of times, the majority rule or any randomized rule will eventually give the expected error rates within each cluster and its overall one equal to their observed ones.

For an error rate of an information flow on serial covariate synergistic feature-groups, the overall error rate is calculated in a weighted fashion. The weighting should be inversely proportion to their individual conditional-entropies. Such simplicity is one significant advantage of adopting an unsupervised learning paradigm. That is, here there is no need to perform the cross-validation as needed in supervised learning paradigms.

\section{Results}
In this section we analyze five simple system data sets from UCI Machine Learning Repository \\ (https://archive.ics.uci.edu/ml/datasets.html). Each data set is chosen for idiosyncratic reasons and characters: 1) the 1st data set with 1D binary response feature is to show why an information flow is more advantageous over Logistic regression model; 2) the 2nd data set with 1D continuous response feature is to recognize the fact that a data set can only sustain limited, not full spectrum, of resolutions of information content as implied by a linear regression model; 3) the 3rd and 4th data sets deals with multiple response features with distinct data types; 4) the 5th data set consists of covariate features of all types: from continuous, discrete to categorical ones, in which all features need to be properly digitally coded.

Here all results of the five data sets presented via information flows are meant to advance our system knowledge with concise and vivid pictorial visualizations. Such an organization of knowledge loci has the potential to take human and machine learning to the next technical level.

\paragraph{$\S$\bf[Brain weight and head size data]}
The first data set consists of two covariate features: 1)brain weight (grams); 2) and head size (cubic cm), for 237 adults classified by two response features: binary gender and age groups: 1) Gender: 1=Male, 2=Female; 2) Age Range: $1=20\sim46$, $2=46^+$, see Gladstone (1905).

An extended version of Logistic regression of gender on brain weight and head size is reported in Fig.1(A) with $\hat{\beta}X_i$ on the horizontal axis. Here $\hat{\beta}$ is the maximum likelihood estimates (MLE) based on Logistic regression model. The evident high degree overlapping between $R_e[1]$ and $R_e[0]$ confirms the inefficiency of Logistic regression on these data. The error rate is $28.3\%$ with threshold at 0.5. This inefficiency due to the artificially imposed homogeneity structure is further contrasted with the a four cluster composition based on the HC tree of $\hat{\beta}X_i$. The three clusters (from the left to the right), except the 4th one, have rather low entropy.

The two possibly-gapped histograms of brain weight and head size are constructed and color-coded with gender-counts into each bins, as shown in Fig.2(A, B), respectively. Each histogram reveals obvious gaps on the two sides of extreme. The heterogeneity manifested through the color-coding and presence of gaps strongly indicates that any homogeneity based on a single distribution assumption is not valid, and more importantly, goes against the true nature of data. Hence such evidence indicates that Logistic regression is not correct for this data set.

The mutual conditional-entropy $E[Y \Leftrightarrow X]$ of the two response features: gender and age, is calculated as being nearly equal to 1. This large entropy value indicates that these two response features represent two separate mechanisms. Thus two separate information flow are reported in Fig.3.

As shown in Fig.3(A), the gender's information flow reveals very evident knowledge loci: a) one extremely small brain weight and head size cluster(C1) is exclusively female; b) an extremely large brain weight and head size cluster(C2) is exclusively male; c) a cluster(C6) of large brain weight and large head size is dominant by male; d) a cluster(C4) of small brain weight and small head size is dominant by female; e) a cluster(C5) median brain weight and head size is mixed. Here we demonstrate that an information flow based on patterned dependency among covariate features can reveal the full spectrum of heterogeneity.

As shown in Fig.3(B), no signs of heterogeneity are seen in the age's information flow, except the cluster(C1) of extremely smallest brain weight and head size. It is clear to see that this information flows can easily adapt to the setting of having more than two age-categories. That is, this information flow platform not only resolves the shortcomings of Logistic regression, but also provides a framework to substitute MNOVA and avoids the required unrealistic distribution assumption, its limitations and ambiguous interpretations altogether.

\begin{figure}[H]
  \centering
   \includegraphics[width=5in]{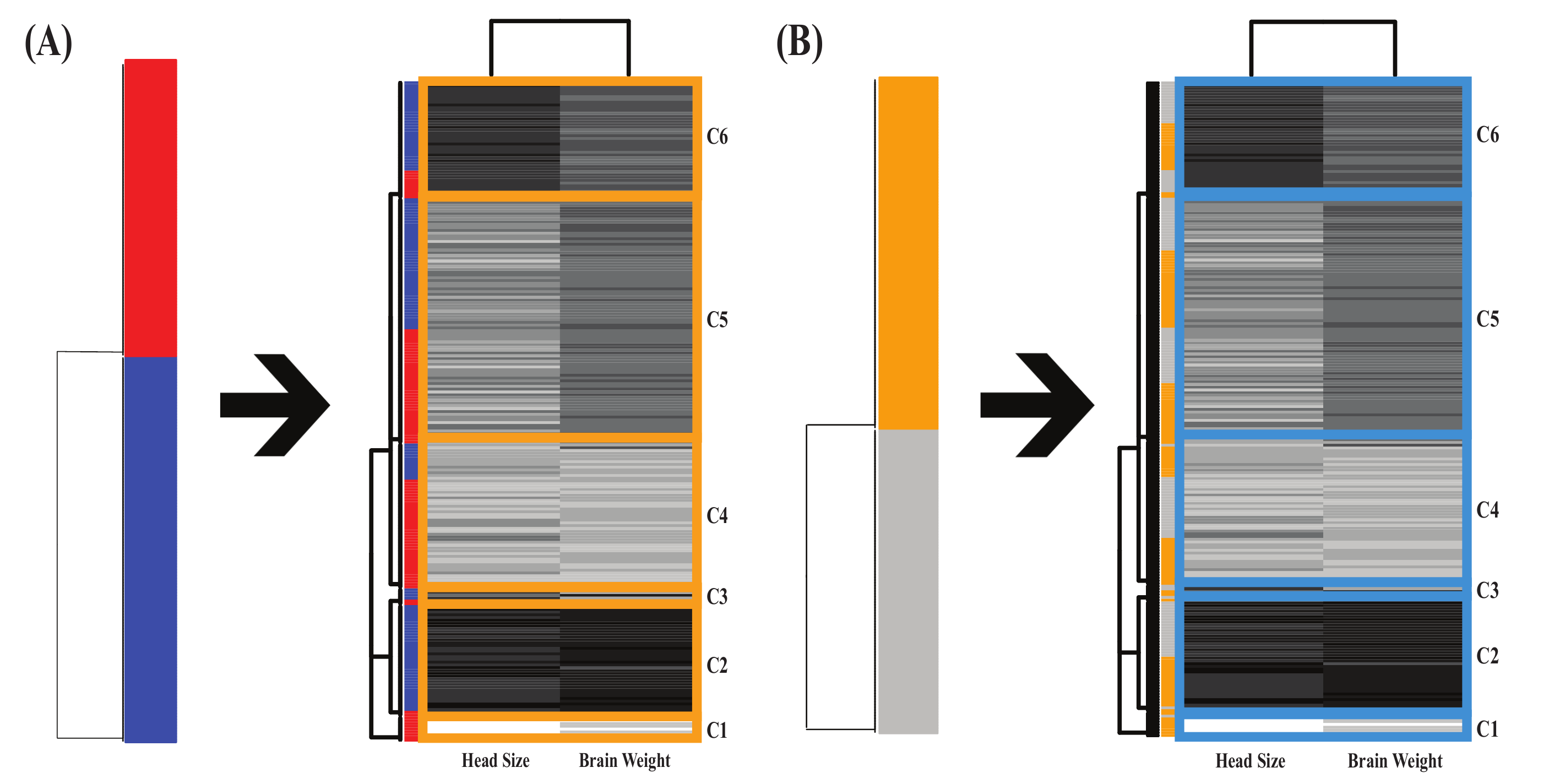}

  \caption{Information flows; (A)for binary-gender; (B)for binary-age. The information flow (A) shows rather evident associative patterns from the gender-tree with male- and female-specific clusters to the DCG-tree based on head size and brain weight with 6 clusters. Except one, all clusters have extremely or relative low entropies. This result shows the effectiveness of information flow over classic logistic regression. The information flows (A) and (B) share a cluster with extreme low values of the two features.}
\end{figure}

One of the original objectives of the investigation as reported in Gladstone (1905) was to obtain a series of reconstruction formulas to predict brain weight given measurement of head size. It is clear that, based on associative patterns of the three features via in the information flow shown in Fig. 3(A), such a prediction would have heterogeneous degrees of precision by taking subject's gender and cluster membership of head size into consideration.

As another demonstration of how dependency structures work, we classify the association between gender and head size into four groups (ranking from all female and extremely small head size to nearly all male and extreme large head size):1) G4; 2) G8; 3) G5-G7; and 4) G1-G3, as shown in panel (F) of Fig.2. Throughout these four groups, the information flow Fig.3(A) demonstrates that females' categorical predictions of brain weight are all correct without ambiguity, while categorical predictions of brain weights for males in the 4th group (G1-G3), who have extremely large head size, can be either extremely heavy or just median heavy. Except such ambiguity on male's prediction, other categorical predictions on are rather precise.

\paragraph{$\S$\bf[Electricity data]}
These data are represent electricity Consumption of 42 provincial town in Great Britain in 1937-1938, see Houthakker (1951). In each town one single response feature was observed: Average total expenditure on electricity, while there were 12 covariate features measured ranging from Average number of consumers(V1), percentage of consumers with two-part tariffs in 1937-38(V2), Average income of consumers(V3), prices on domestic tariffs in 1933-34(V4), 1935-36(V5), 1937-38(V6), Marginal price of gas 1935-36(V7), 1937-1938(V8), Average holdings of heavy electric equipment bought (V10) and per two-part consumer consumption 1937-38(V10), 1935-36(V11) and 1933-34(V12). Here we report information flows according to two scales of clustering on the response feature: one fine-scale (with 5 clusters and one extreme outlier) and one coarse-scale (with two clusters) clustering compositions of the response feature, as shown in Fig.4(A,B) respectively.

The heatmap of mutual conditional-entropy of 12 covariate features, which is superimposed by a DCG tree, shows four synergistic feature groups in Fig.5(A). Three information flows from response's fine-scale perspective are reported in three panels in Fig.5(B,C,D). The information flow from the response to the synergistic feature-group$\#2$ (V2, V3, V10-V12), as shown in Fig.5(B), demonstrates that each branch of the three-cluster level of DCG tree ${\cal T}[{\cal M}^{(Co)}_{\circledR,2}]_R$ is coupled with rather clear dependency structures marked by uniform and evident block patterns in the heatmap.

Though each of these three cluster indeed consists of mixed color-coded memberships of the three response's clusters, two out of three of them are significantly non-random. Their observed conditional-entropies are calculated (with P-values in parenthesis) as $(0.65(0.0),1.39(0.38),1.05(0.004))$ (from top to bottom on the 3-cluster level) in relative to the entropy on the response side calculated as 1.62. The p-values are evaluated through the simulation scheme of simple random sampling without replacement on the subject space with respect to the response's 6 color-coding. Hence we conclude that the presences of relative low entropy-values with extremely low p-values strongly indicate that response-to-covariate associative patterns are evident, but not exclusive. 

Similar conclusions can be made for the other two information flows: 1) one union of $\# 1$ and $\# 2$ synergistic feature groups having 8 features, as shown in Fig.5(C); 2) and one union of $\# 1$, $\# 2$ and $\# 3$ synergistic feature groups having 10 features, as shown in Fig.5(D).

It is important, but not difficult to see that such non-exclusiveness in the middle covariate cluster of information flow in Fig.5(B), the bottom one in Fig.5(C) and the top one in Fig,5(D), is primarily due to the presence of three response categories with relative large values. This fact critically points out that this data set can not sustain such a fine resolution on the response feature. Hence we conclude that overall the fine scale structure with 6 clusters chosen for the response feature is supported only in part. In other words, this data set can not afford such a fine scale separation on response features. How about the coarse-scale one?

Two information flows from response's coarse-scale perspective are reported in Fig.6. Through the DCG tree ${\cal T}[{\cal M}^{(Co)}_{\circledR}]_R$ (including all 12 features) superimposed upon a block-patterned covariate matrix, the information flow, as shown in Fig. 6(A), reveals four major clusters are coupled with clear block patterns. Three of them have zero conditional-entropies by having exclusive memberships belonging to one of the two response clusters. However the fourth one is a mixed.

In contrast the second information flow, as shown in Fig. 6(B), the DCG tree ${\cal T}[{\cal M}^{(Co)}_{\circledR,2}]_R$ based on $\# 2$ synergistic feature group pertaining to the first of the serial heatmaps on the right reveals a tree level with three clusters: 1) two  exclusively contains members from the cluster of small-value cluster (in black) of response; 2) one is nearly exclusively dominated by members of the cluster of large-responses. That is, the exclusiveness of the linkage between response on coarse scale and covariate on three cluster scale is established.

Correspondingly two aspects of system understandings are derived as follows. The first aspect is that the small-value cluster of response contains heterogeneity caused by two block-patterns of covariate features: a) extremely large $V2$-value and extremely small $(V3, V10-V12)$-values; b) median $V2$-value and median $(V3, V10-V12)$-values. The second aspect is that the large-value cluster of response is attributed to extremely small $V2$-value and extremely large $(V3, V10-V12)$-values. These two aspects spell out the first important parts of knowledge loci based on the clear dependency structures of synergistic covariate feature-Group $\#2$. More knowledge loci are available along the 2nd through 4th heatmaps of this information flow. These knowledge loci can be used to further correct, or at least update and improve the misclassifications made in the 1st heatmap as follows.

Predictions are made and conformed via ``majority rule'' within each cluster identified across different heatmaps on the right. As illustrated in Fig.6(B), as if those number-marked subjects were missing their response feature measurements, then each heatmap gives rise to a set of predicted values. A final decision for each individual would be reached by simply conducting weighted averaging of the four predicted values with weights inversely proportional to the four corresponding conditional-entropy values. This is an error-correcting mechanism provided by using information flow with serial DM-computed heatmaps.

At the end of this example, it is strongly emphasized that the information flow in Fig. 6(B) is much more proper and informative than the one in Fig. 6(A) because the four synergistic covariate feature groups are somehow antagonistic to each other. Therefore a platform for them to show their idiosyncratic dependency is needed. The information flow is designed to provide such a platform.

\begin{figure}
 \centering
 \includegraphics[width=5in]{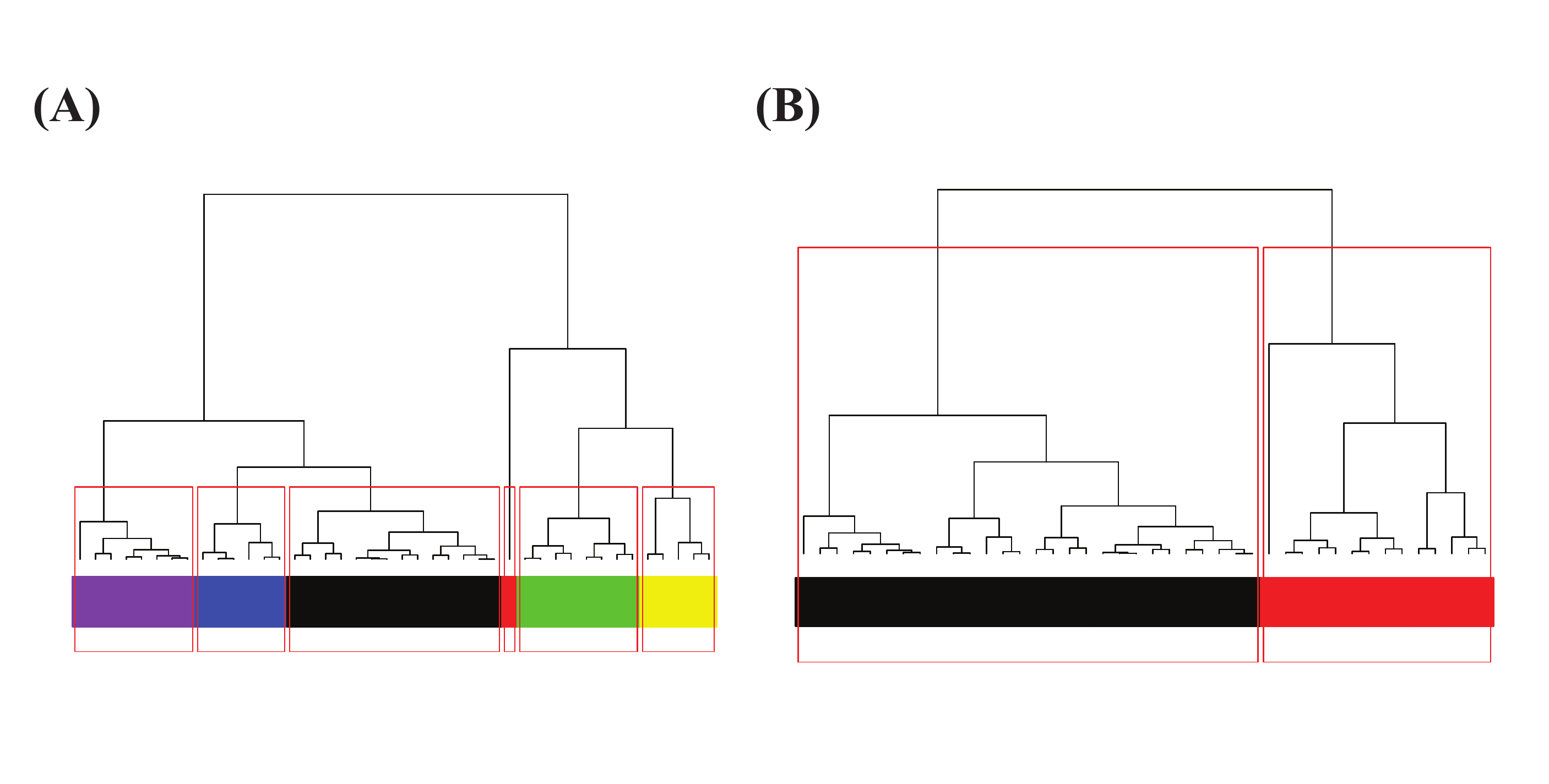}

 \caption{Response Hierarchical clustering trees: (A) the fine-scale with 6 color-coded clusters; (B)the coarse-scale 2 color-coded clusters. It is intuitive that the task of successfully differentiating among the 6 fine-scale clusters of (A) would need much more covariate information than the task of differentiating between the two coarse-scale clusters of (B).}
 \end{figure}

\begin{figure}
 \centering
  \includegraphics[width=5in]{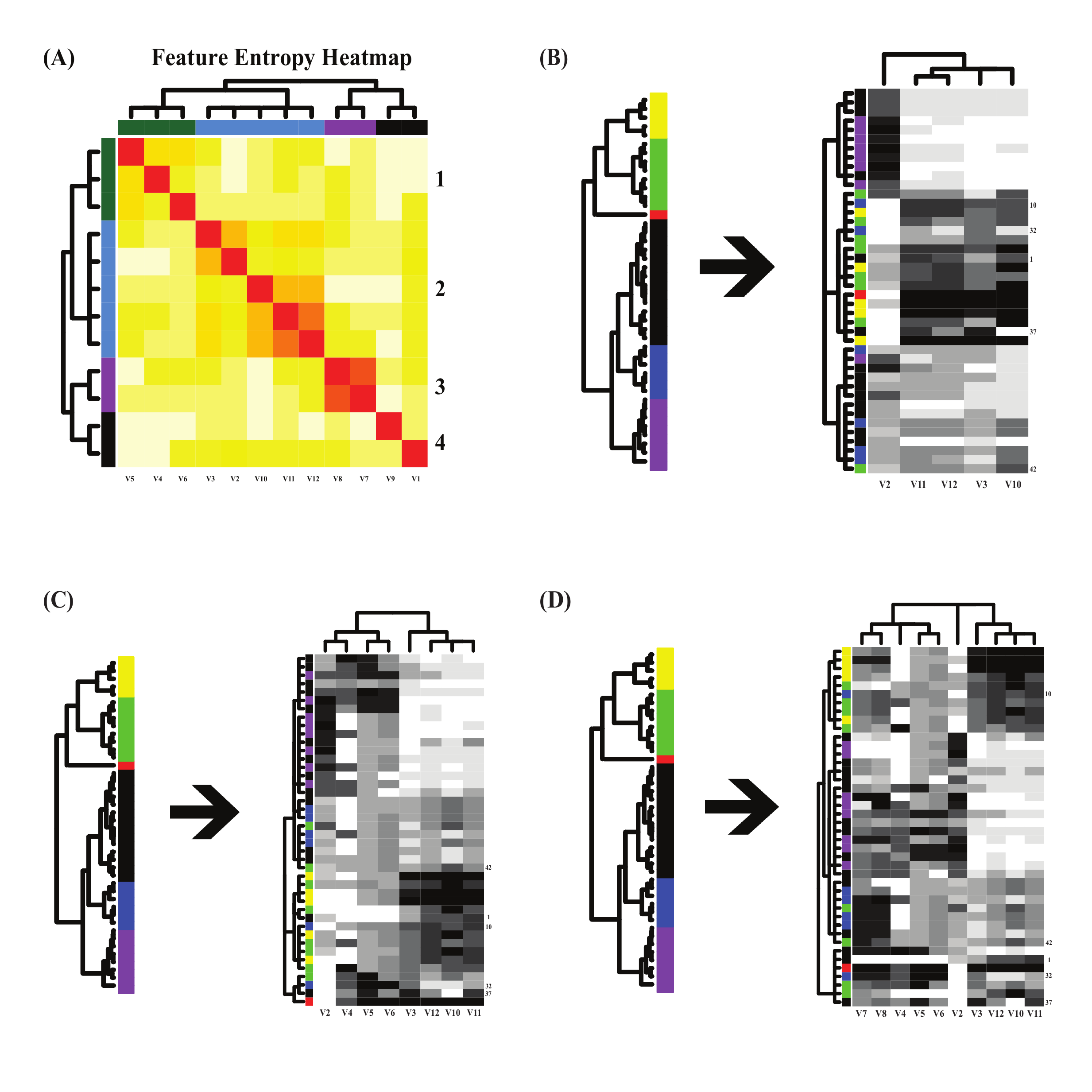}

 \caption{Information flows from response's fine-scale perspective: (A) Mutual conditional-entropy matrix superimposed with DCG tree with 4 synergistic feature-groups; The information flows from the response to (B) $\# 2$ synergistic feature-group; (C) $\# 1\& \#2$; (D) $\#1\& \#2 \& \#3$ synergistic feature-groups. The misclassified subjects' ID numbers are attached to the right side of each heatmap.}
 \end{figure}

\begin{figure}
 \centering
  \includegraphics[width=5in]{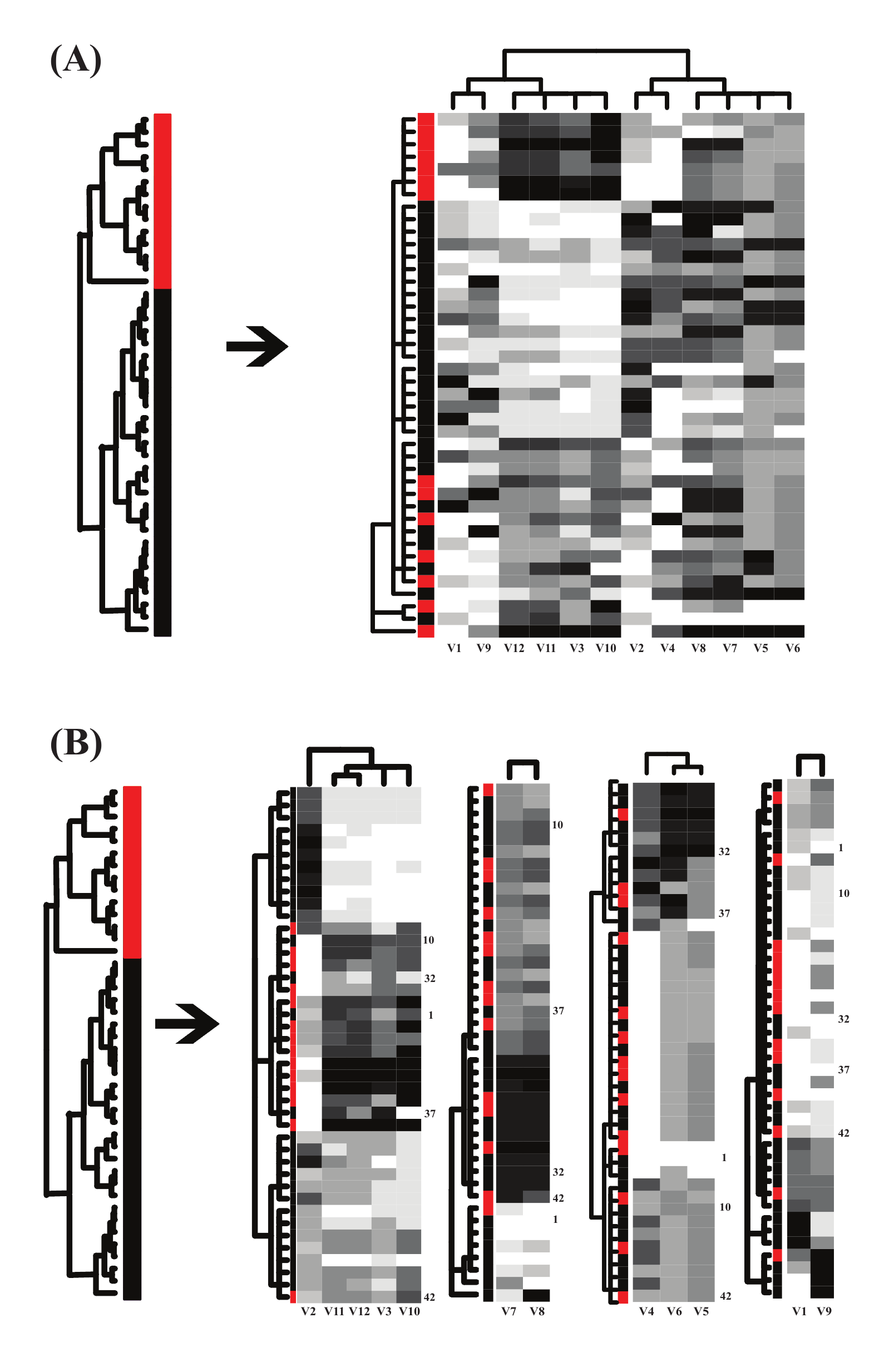} 

 \caption{Information flows from response's coarse-scale perspective. The information flows from the response to (A) $\# 1\& \#2 \& \#3 \& \#4$ synergistic feature-groups ; (B) serial $\#2$, $\#3$, $\#1$ and then $\#4$ synergistic feature-groups.}
 \end{figure}

The original goal of this example according to Houthakker (1951) is two-fold: analyzing electricity demand and investigating monthly fluctuations. The hope was to incorporate results from these two parts to achieve a comprehensive study of the various features on electricity consumption. Yet this goal could not be realized in the original study due to insufficient information from the data as stated by the author. However, on the coarse scale of the response feature here, this goal can be achieved via our information flows. They indeed provide very comprehensive system understanding on the electricity consumption during the two year period from the 42 provincial towns in Britain.

\paragraph{$\S$\bf[Patterns of Bird Species in Andes Mountains]}
In Ecology during 1970s, biogeographers assumed that continental biota found on high mountain tops are as isolated from one another as true islands. In order to test whether high mountain biota have insular distribution patterns, Vuilleumier(1970) collected data of bird species among ``island'' of mountain tops in 15 regions of the p$\acute{a}$ramo vegetation in the Andes of Venezuela, Colombia and northern Ecuador.

There are 3 response features: Total Number of Species(V2), Number of species of South American origin(V3) and Number of endemic taxa(V4), and 7 covariate features: Area(V6), Base altitude(V7), Elevation(V8), Distance from Paramo(V9), Distance to nearest island of vegetation(V10), Distance to nearest island in south(V11) and Distance to nearest large island(V12), see details in Vuilleumier(1970).

Two mutual conditional-entropy matrices for the response and covariate features are separately computed, as shown in Fig.7 (A,B). The response's heatmap on left hand side of Fig.7 (C) from DM clearly shows two patterned blocks  that indicates strong joint dependency: largeness-vs-smallness, among response features. In contrast, the covariate's heatmap on the right hand side of Fig.7(C) also clearly shows the joint dependency of  7-dim covariates in two scales: 1) two patterned blocks; 2) each block is intricately divided into two sub-blocks.

The first information flow linking the structural dependency on both sides, as shown in Fig.7(C), reveals a perfect linkage of heterogeneity from the response's two blocks to the covariate's 4 sub-blocks. This resulted perfect linkage of heterogeneity is surprising in the sense that the largeness-vs-smallness of response features is determined by rather intricate differences between sub-block belonging to each of the two block of covariate features. The second information flow in Fig.7(D) from 2-dim responses (V2 and V3) to a couple of 3-dim and 4-dim covariates also reveals the same kind of heterogeneity as clear as the first one.

\begin{figure}[H]
  \centering
    \includegraphics[width=5in]{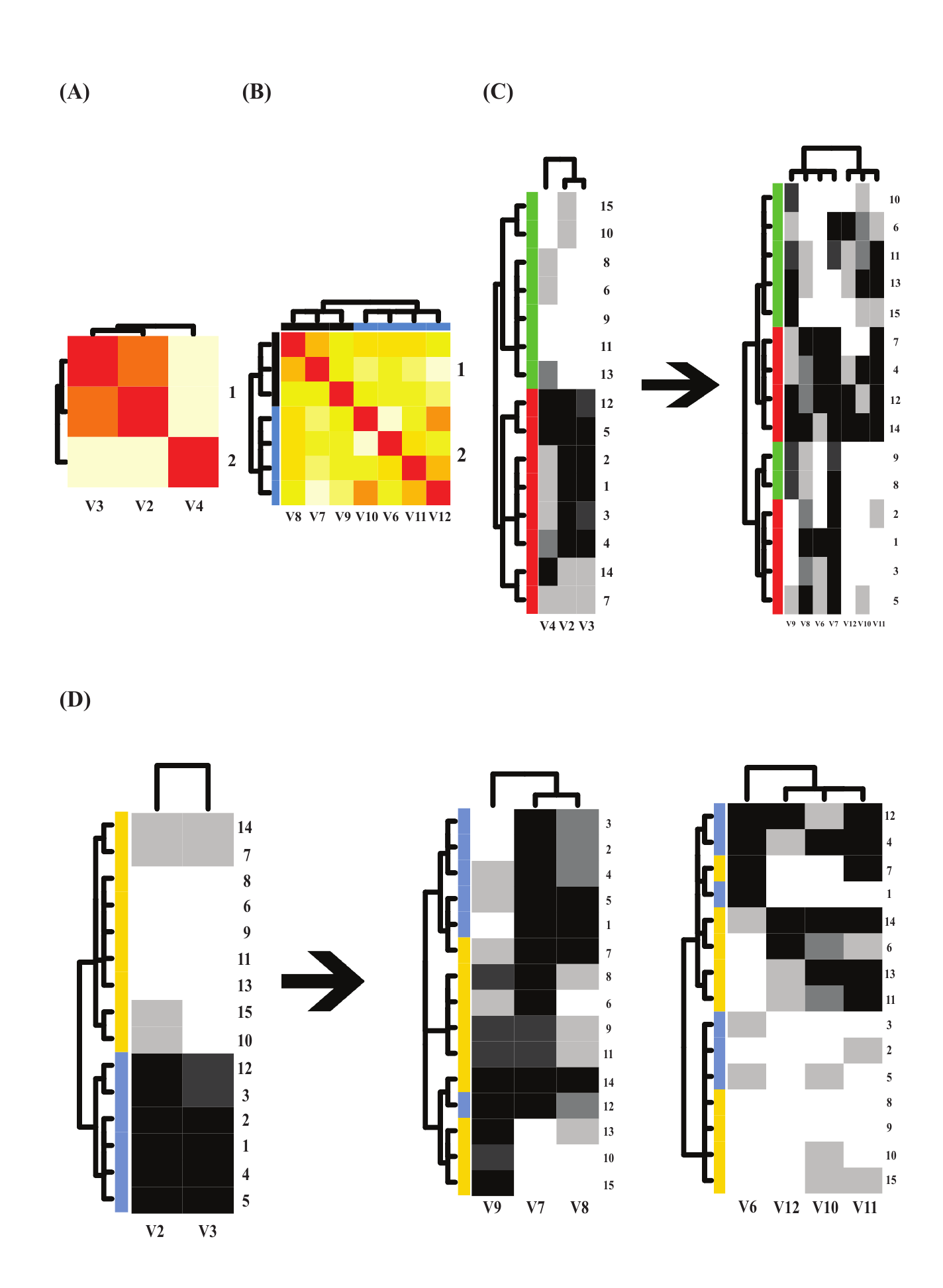}

\caption{Mutual conditional entropy matrices and two information flows on bird data. (A) Mutual conditional entropy matrix of 3 response features divided into two synergistic groups; (B) $7\times 7$ mutual conditional entropy matrix of covariate features with two color-coded synergistic groups; (C) Information flow from the response heatmap to the covariate heatmap showing heterogeneity; (D) Information flow from two response features v2 and V3 to two covariate heatmaps pertaining to the two synergistic groups.}

\end{figure}

Such splitting heterogeneity seen in the information flows conclude that the clearly bifurcated linkages from the response features to the covariate features strongly indicate that the high order dependency among covariate features is the driving forces underlying this biogeographic system, on one hand. On the other hand, it interestingly undermines the linear regression reported in the original paper.

The original investigation in Vuilleumier(1970) employed stepwise linear regression of one response feature at a time and reported very well statistical modeling fitting. Here we like to point out the fact that such statistical results likely were caused by over-fitting. The a linear hyper-plan based on 7 covariate features can easily over-fit the small number $(15)$ of data points.

This example very well demonstrates the essence and importance of computing joint dependency among response and covariate features in order to discover evident heterogeneity through an information flow as shown in the Fig.7(C). Thus it is worth emphasizing that the knowledge loci contained in this data are organized on the fine, not coarse, scale of block patterns. Since heterogeneity hardly can be accommodated by homogeneous linearity as assumed in the regression model, the results of linear regression analysis become misleading and dubious.

\paragraph{$\S$\bf[Height and Various Stature Measurements Data]}
The fourth data set is consisting of 33 female police-department applicants. Each applicant has her standing Height(V2) and sitting height(V3) measured as two response features, and upper arm length (V4), forearm (V5), hand (V6), upper leg (V7), lower leg (V8), foot (V9), forearm/upper arm (V10), lower leg/upper leg (V11) are measured as seven covariate features, see Lafi and Kaneene(1992). Mutual conditional-entropies on response and covariate sides are computed and shown in Fig.8 (A) and (B), respectively. Two synergistic covariate feature-groups are identified:$ Group \# 1=\{V4, V5, V6, V7, V8\}$ and $ Group \# 2=\{V9, V10, V11\}$ .

The response's heatmap resulted from DM, as shown on the left hand side of Fig.8(C), clearly reveals three clusters coupled with evident block patterns. Thus it is not only reasonable, but necessary to take both two dimensional features as response simultaneously. One information flow from the 2-dim response features to $\#1 \& \#2$ covariate features groups is carried out and reported in Fig.8(C).  Also two major and one small covariate clusters are also supported by clear block patterns. One major covariate cluster is dominated by one response cluster members (Green color-coded) with a few coming from the the other two response clusters. We perform simple random sampling without replacement similar to permutation test to conform this pattern formation. The observed entropy is relatively small $0.87$ comparing with 1.09 the overall entropy from response with the p-value $0.01$. The other major covariate cluster primarily has mixed memberships of two response clusters (blue and orange color-coded). The observed entropy is $0.84$ with its p-value $0.008$.

The second information flow from the 2-dim response features to a serial of $\#1$ and then $\#2$ covariate feature-groups is shown in Fig.9(d). The first heatmap on the right hand side of information flow reveals two covariate clusters. These two clusters have mixed memberships of three response clusters like the manifestation in the first information flow. The observed entropies with their p-value in parenthesis are calculated as $0.94(0.008)$ (for green-dominant one) and $0.89(0.019)$ (for the mixture of blue and orange), respectively.

Through pattern confirmations with small p-values are resulted in both two information flows, the 2nd information flow clearly indicates that extreme small standing and sitting heights are associated particularly with small values of features belonging to $\#1$ feature-group; in contrast large standing or sitting heights are highly associated large values of the same five features in $\#1$ group. Again we demonstrate that these knowledge loci are displayed through the linkages between response's and covariate's dependency structures. This and the above examples nicely illustrate the exploratory nature of our proposed categorical pattern matching and the resolutions to the issue of multiple response.

\begin{figure}[H]
  \centering
    \includegraphics[width=5in]{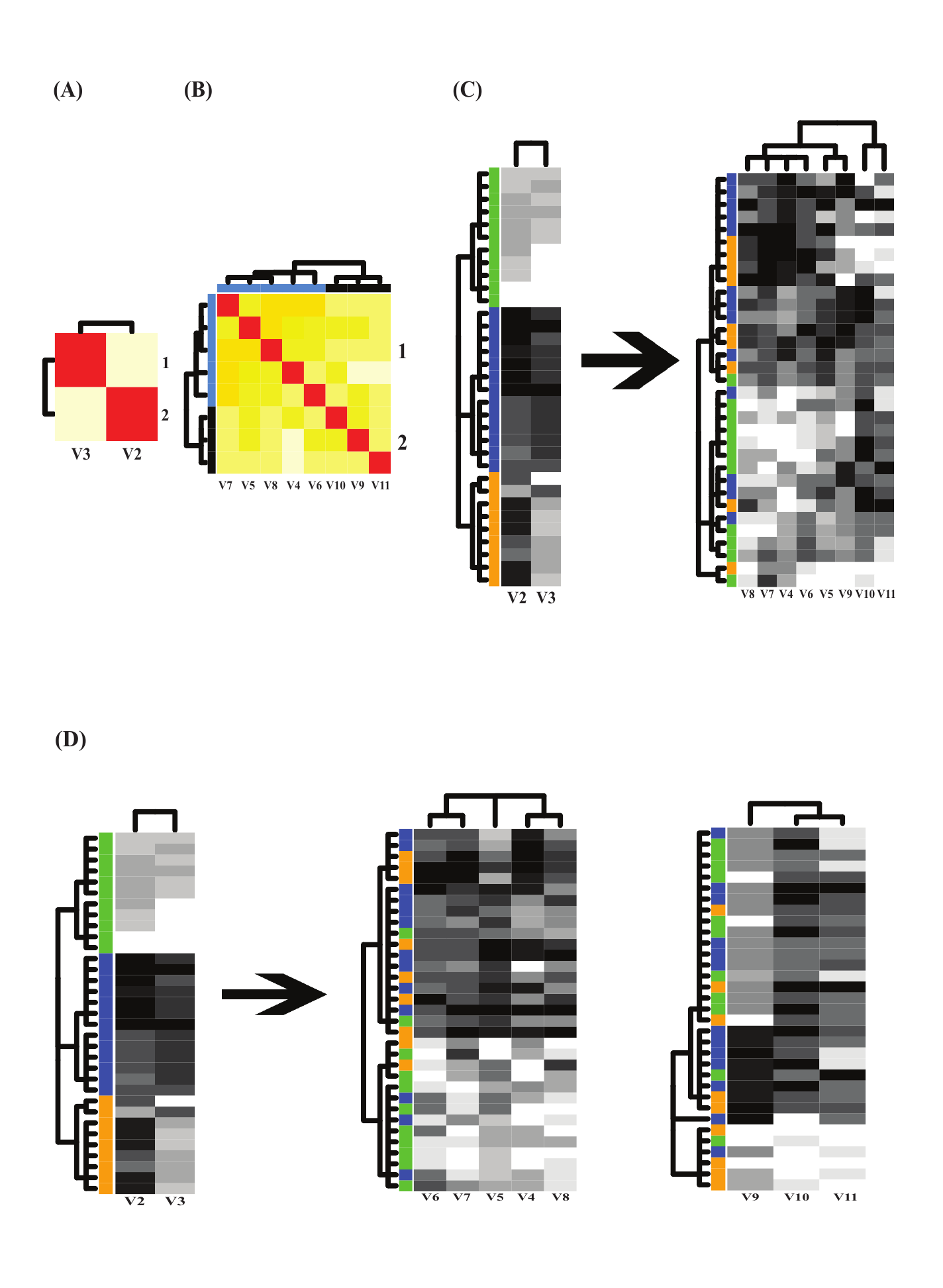}

\caption{Information flow of Height data. (A) and (B) for the mutual conditional-entropy matrices for response  and covariate features; (C) Information flow to all covariate features; (D) Information flow to $\#1$ feature-group and then $\#2$ feature-group.}
\end{figure}

The original investigation in Lafi and Kaneene(1992) was concerned about the issues arising from multicollinearity among the 8 covariate features, including the sitting height, in linear regression analysis with the standing height as the response feature. Again it is rather unnatural that two highly related features: sitting and standing heights, as seen in the Fig. 8(A), are separated by the divide between response and covariate. This certainly was done due to the fact of lacking statistical methodology for accommodating Multiple response.

The principle component analysis (PCA) was used to convert the 8 features into a few independent ``factors'' to alleviate effects of multicollinearity. Again such linearity based artificial factors made the regression results very hard for interpretation. In contrast, our information flow clearly and naturally reveals patterns of response features and links them with associative patterns based on groups of synergistic covariate features with evident heterogeneity.

\paragraph{$\S$\bf[Heart disease]}

This dataset taken from UCI machine learning repository contains 13 features and 270 human subjects. Among 13 features, there are 5 continuous ones: Age(V1), Resting Blood Pressure(V4), Serum Cholestorol(V5), Maximum Heart Rate Achieved(V8),Oldpeak(V10); 3 binary Variables : Sex(V2), Fasting Blood Sugar(V6)($>$ 120 mg/dl), Exercise Induced Angina(V9); and 5 categorical ones:
Chest Pain Type(V3, with values 1,2 3,4), Resting Electrocardiographic results (V7, with values 0,1,2), Slope of the Peak Exercise ST segment(V11, with 1: upsloping; 2:flat; 3:downsloping), Number of Major Vessels colored by Fluoroscopy(V12, with 4 values from 0 to 3), thal (V13, with 3 = normal; 6 = fixed defect; 7 = reversable defect), see Detrano, et al. (1989) for details. This example illustrates how to handle digital coding for mixed data types.

Each binary and categorical features are digitally coded for making the digital coding more comparable with continuous features. The coding scheme for a categorical one is based on its closest non-categorical feature.
\noindent [Digital Coding for binary and categorical features:]
\begin{description}
\item[1. Binary:] $\{0\}\rightarrow 0$ and $\{1\}\rightarrow 5$;
\item[2. Categorical-V3:] being close to binary V9, $\{1, 2, 3\}\rightarrow 0$; $\{4\}\rightarrow 5$ ;
\item[3. Categorical-V7:] being close to continuous V8, $\{0\}\rightarrow 9$; $\{1\}\rightarrow 3$; $\{2\}\rightarrow 7$;
\item[4. Categorical-V11:] keep ordinal order with $\{1\}\rightarrow 3$; $\{2\}\rightarrow 6$; $\{3\}\rightarrow 9$;
\item[5. Categorical-V12:] keep ordinal order with $\{0\} \rightarrow 0 $; $\{1\}\rightarrow 3$; $\{2\}\rightarrow 6$; $\{3\}\rightarrow 9$;
\item[6. Categorical-V13:] being close to binary V2, $\{3\}\rightarrow 0$; $\{6, 7\}\rightarrow 5$.
\end{description}

The mutual conditional-entropy matrix shows two synergistic feature groups in Fig.9(A). The heatmap of involving all covariate features, as shown in Fig.9(B), reveals their joint dependency via two scales of pattered blocks: 1) the fine scale having 9 clusters, denoted by G1 through G9; 2) the coarse scale having 3 conglomerate clusters, (G1, G2), (G3, G4) and (G5, G6, G7,G8, G9).
The information flow from response's patient and healthy subject clusters to involving all covariate features, as shown in Fig.9(B), discovers high degrees of heterogeneity of covariate patterns within the patient as well as healthy subject clusters. It is noted that we also explore information flows based on either of the two synergistic feature-groups. They are not as effective as the one involving with all covariate features.

Further, via simple random sampling without replacement scheme, the classification performance pertaining to two scales of clustering compositions: 3 clusters (Yellow color-coded boxes) and 9 clusters (Black color-coded bars), in the information flow are evaluated through 1000 simulations and presented in box-plots of $95\%$, as shown in Fig.10. We see that observed entropies (Blue for 3-cluster scale and Red for 9-cluster scale) below their corresponding boxes indicate significant results, that is, the clusters with non-random compositions of patients and healthy subjects with P-values less than $5\%$. It is noted that a smaller cluster size would render a longer $95\%$ box.

\begin{figure}
 \centering
   \includegraphics[width=5in]{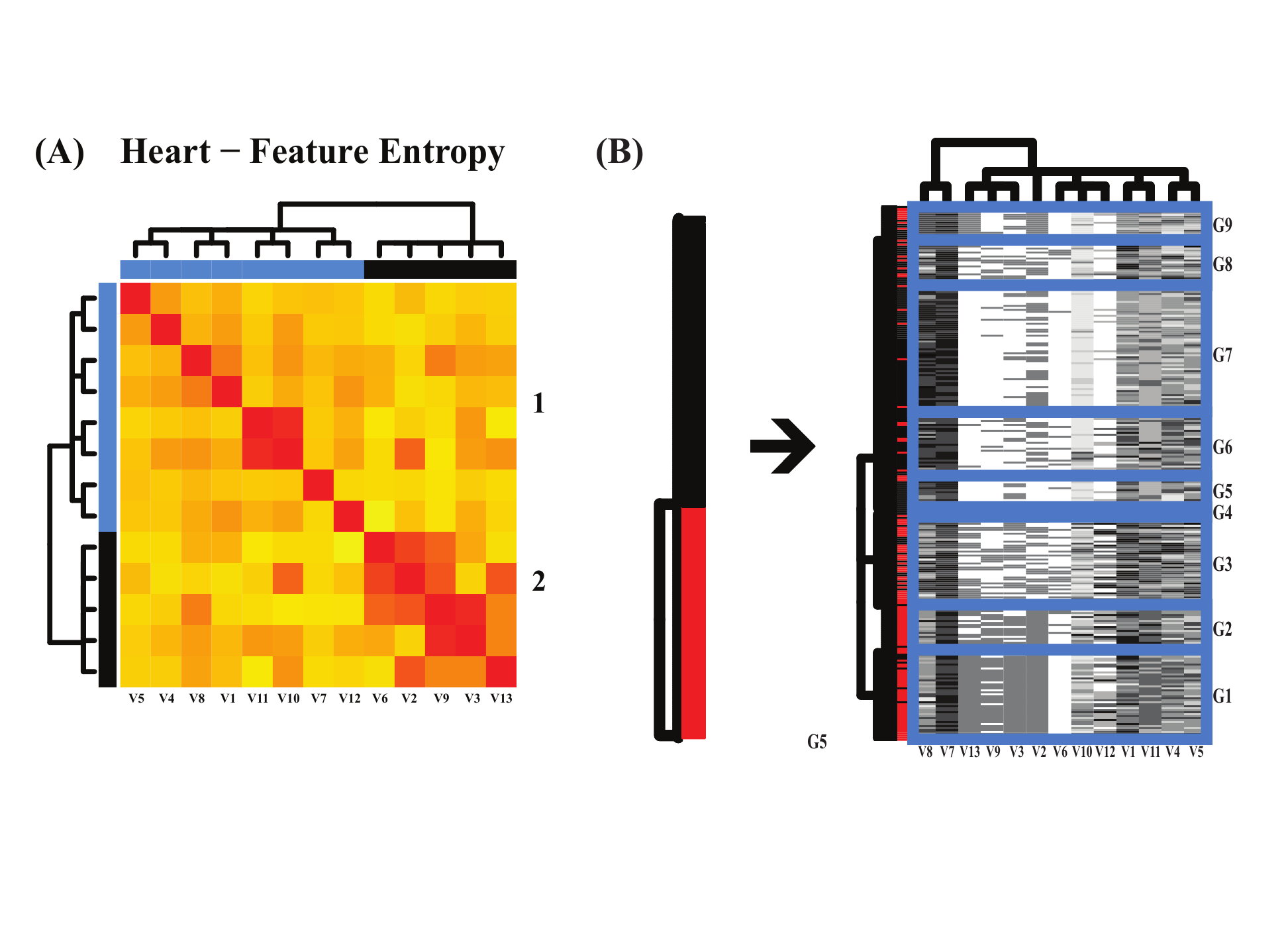}

 \caption{Heatmaps via DM on heart disease data: (a) Mutual entropy matrix of all features with two synergistic groups; (b) Coupling geometries of all features. Red color for patients, Black for healthy subjects.}
 \end{figure}

 \begin{figure}
 \centering
   \includegraphics[width=5in]{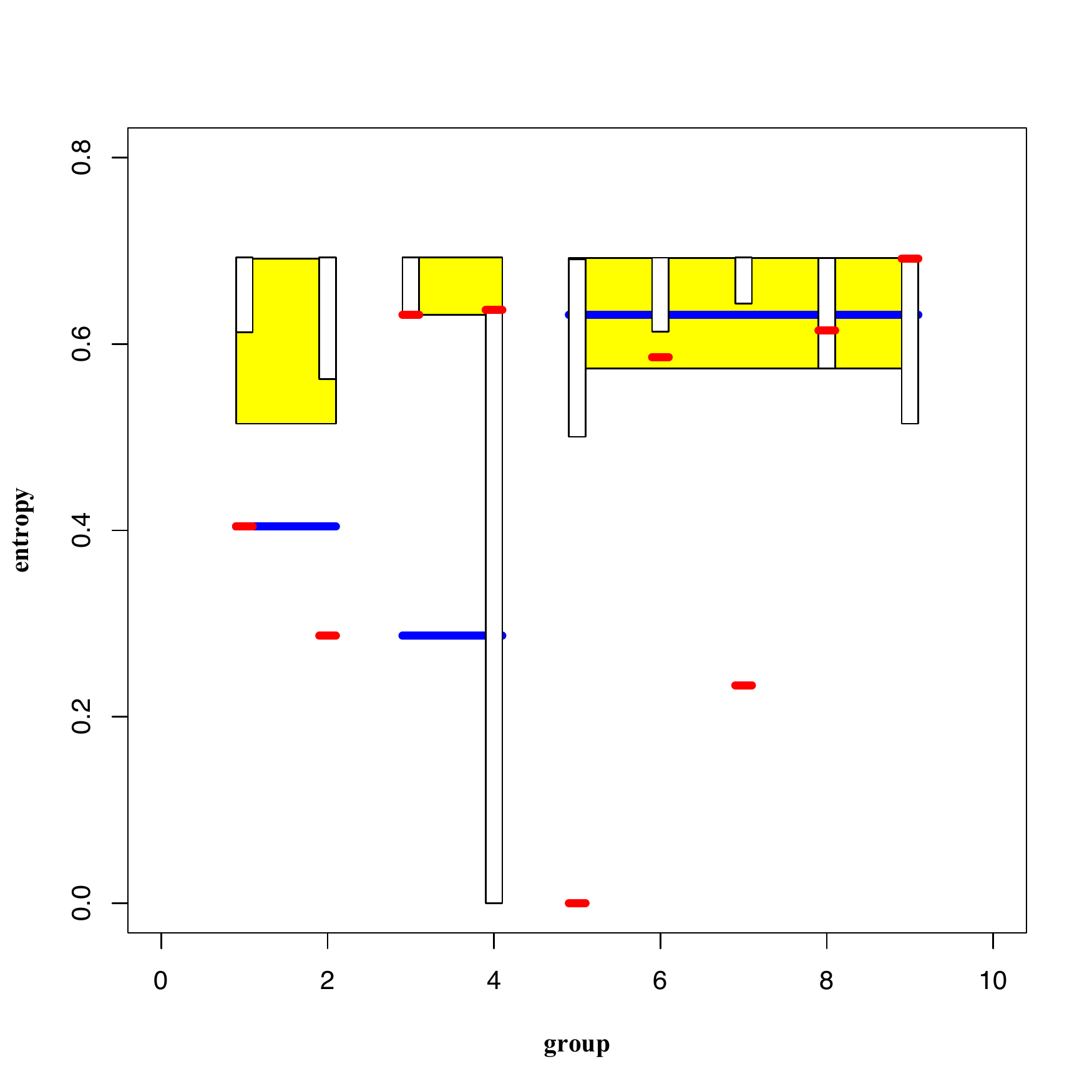}

 \caption{Classification performances of an information flow on two scales: $95\%$ Box-plots of the three-cluster scales is in Yellow color with observed entropy being marked in Blue, while the nine-cluster scale one in Black with observed entropies being marked in Red. The clusters from the left-to-right are arranged exactly to correspond to clusters from bottom-to-top in Fig.9(B). Each box is built based on 1000 simulated entropy values via simple random sampling without replacements.}
 \end{figure}

\section{Discussion}
In this paper we develop one unified platform for discovering system knowledge from data analysis. The system knowledge loci are organized and represented through one single or multiple information flows. An information flow represents matrix-based causal linkages from response's computed structural dependency to covariate's computed patterned dependency. The patterned dependency computed via Data Mechanics on a matrix is revealed through multiscale blocks framed by a clustering tree superimposed on a group of synergistic features and another clustering tree on the common space of system subject. That is, such patterned dependency summarizes essential information contents on response and covariate data matrices, respectively, without involving potential distortions possibly caused by unrealistic modeling or distribution assumptions. Upon this platform for categorical-pattern-matching, such linkages between the two computed structural dependency is capable of bringing out readable and visible system knowledge and understandings. We are confident that such an information flow can make data-driven computing and learning coherent and fruitful in discoveries in sciences.

It is worth re-emphasizing that resolving scientific problems by extracting authentic information from data is the primary goal of data analysis.  Modeling a complex system with a set of convoluted mathematical functions and equations is an activity belonging to the category of synthesis. It is expected all but impossible when the amounts of data are really large. Since a man-made model hardly has capacity of accommodating natural multiscale heterogeneity and dependency embedded within data. Unfortunately most of statistical modelings and methodologies fail with respect to such criteria.

Here we mention and discuss the narrow perspective tied to any model selection techniques in statistics. Beside the ad hoc criterion too narrowly focused on detailed variations, like sum of squared error (SSE), a model-selection technique would choose only one set of covariate features from a fixed ensemble of potential models. How about another set of covariate features, which results in a just sightly larger SSE than the minimum one? It seems like that no extra information can be offered from this second best set of covariate features. This is totally not true because there might exist several distinct and meaningful mechanisms simultaneously associating with one single dimensional response feature. For instance, consider a study on causes of obesity. Wouldn't the potential causes of obesity become more and more complex when more than more subjects are included into the study? More subjects certainly will bring in more diverse and different psychological factors, environmental conditions, cultures and genetic makeups and many others. They are going to tightly tangle together. So there are many causes. Not just One. 

Further all model selection techniques assume an implicit fundamental assumption that the ensemble of potential models is invariant with respect to the number of involving subjects. Like the conditioning argument in all regression analysis, this invariance assumption is another strong evidence of ignorance of structural dependency on the covariate side. To be more specific, as the ensemble of observed system subjects becoming larger, its subject's ``community structure'' is more fully exposed as in the above obesity study. That is, distinctions and gaps among these potential multiscale communities are more evidently expressed through block patterns due to large as well as finer scales dependency among covariate features. The presence of finer and finer scales dependency structures is exactly behind the explanation of the fact no models are correct when the sample size is really big. But this invariance assumption imposed by all model selection techniques strictly require practitioners to blindly give up the truth that there are multiple mechanisms are behind whatever trends of one single response feature. On the other hand, this critically unreasonable assumption of fixed ensemble of potential models with respect to sample size also reflects the impossibility of the issue of how to properly grow the ensemble as sample size increase.

At the end, we remark that our matrix-based causal linkage approach can also fundamentally resolve the recent issue of reproducibility of research results for publications in major scientific journals. The reason is that even though this reproducibility concern has pressured scientists to be more vigilant and rigorous when they conduct and report their data analysis, unintentional or careless mistakes or human fallacies can still creep in the modeling and affect summarizing parameter values. Requirement of submitting the original data in the submission process would prevent potential human errors to some extent. However effects of man-made assumptions, particularly involved in complicate modelings, are still hard to be filtered out and prevented from contributing to, or even dominating the implications via reported statistical results.

\section{references}
\begin{description}
\item[]Anderson, P. W. (1972). More is different. Science 177, 393-396 .
\item[]Tukey, John W. (1962) The Future of Data Analysis.  The Institute of Mathematical Statistics, Ann. Math. Statist, 33(1), 1-67.
\item[]Detrano, R., Janosi, A., Steinbrunn, W., Pfisterer, M., Schmid, J..J., Sandhu, S., Guppy, K. H., Lee, S. and Froelicher, V.(1989) International application of a new probability algorithm for the diagnosis of coronary artery disease. Am. J. Cardiol. 64, 304-310.
\item[]Hsieh Fushing and McAssey, P. M. (2010) Time, temperature and data cloud geometry.  Physics Review E, 82, 061110-10.

\item[]Hsieh Fushing, Chen, C., Liu, S.-Y. and Koehl, P. (2014). Bootstrapping on undirected binary network via statistical mechanics. J. of Statistical Physics, 156, 823-842.
\item[]Hsieh Fushing and Chen, C. (2014).  Data mechanics and coupling geometry on binary bipartite network. PLoS One, 9(8): e106154. doi:10.1371/journal.pone. 0106154.
\item[]Hsieh Fushing, Hsueh C-H, Heitkamp C, Matthews M., Koehl P. (2015) Unravelling the geometry of data matrices: effects of water stress regimes on winemaking. Journal Royal Society- Interface 20150753. http://dx.doi.org/10.1098/rsif.2015.0753
\item[]Hsieh Fushing and Roy, T. (2017). Complexity of Possibly-gapped Histogram and Analysis of Histogram (ANOHT). arXiv:1702.05879v1[Stat.ME] 20Feb2017.
\item[]Gladstone, R. J. (1905) A Study of the Relations of the Brain to the Size of the Head. Biometrika, 4,105-123.
\item[]Houthakker, H. S.(1951) Some Calculations on Electricity Consumption in Great Britain. Journal of the Royal Statistical Society. Series A (General), 114, 359-371.
\item[]Janyes, E. T. (1957).  Information theory and statistical mechanics,106 (4): 620–630.
\item[]Kolmogorov, A. N. (1965). Three approaches to the quantitative definition of information. Problemy Peredachi Informatsii. 1, 3-11.
\item[]Lafi,S.Q. and Kaneene, J.B. (1992). An explanation of the use of principal-components analysis to detect and correct for multicollinearity, Preventive Veterinary Medicine, 13, 261-275
\item[]Shannon, C. E. (1948). A Mathematical Theory of Communication. Bell System Technical Journal. 27 (3): 379–423.
\item[]Shannon, C. E. (1951). Prediction and Entropy of Printed English. Bell System Technical Journal. 30 (1): 50–64.
\item[]Vuilleumier, F. (1970). Insular Biogeography in Continental Regions. I. The Northern Andes of South America. The American Naturalist, 104, 373-388.
\end{description}

\end{document}